\documentclass[preprintnumbers,amsmath,amssymbm,prd]{revtex4}
\usepackage{epsfig}
\usepackage{graphicx}

\begin{document}
\title{The spinning Kerr-black-hole-mirror bomb: A lower bound on the radius of the reflecting mirror}
\author{Shahar Hod}
\affiliation{The Ruppin Academic Center, Emeq Hefer 40250, Israel}
\affiliation{ } \affiliation{The Hadassah Institute, Jerusalem
91010, Israel}
\date{\today}

\begin{abstract}
\ \ \ The intriguing superradiant amplification phenomenon allows an
orbiting scalar field to extract rotational energy from a spinning
Kerr black hole. Interestingly, the energy extraction rate can grow
exponentially in time if the black-hole-field system is placed
inside a reflecting mirror which prevents the field from radiating
its energy to infinity. This composed
Kerr-black-hole-scalar-field-mirror system, first designed by Press
and Teukolsky, has attracted the attention of physicists over the
last four decades. Previous numerical studies of this spinning {\it
black-hole bomb} have revealed the interesting fact that the
superradiant instability shuts down if the reflecting mirror is
placed too close to the black-hole horizon. In the present study we
use analytical techniques to explore the superradiant instability
regime of this composed
Kerr-black-hole-linearized-scalar-field-mirror system. In
particular, it is proved that the lower bound
${{r_{\text{m}}}\over{r_+}}>{1\over
2}\Big(\sqrt{1+{{8M}\over{r_-}}}-1\Big)$ provides a necessary
condition for the development of the exponentially growing
superradiant instabilities in this composed physical system, where
$r_{\text{m}}$ is the radius of the confining mirror and $r_{\pm}$
are the horizon radii of the spinning Kerr black hole. We further
show that, in the linearized regime, this {\it analytically} derived
lower bound on the radius of the confining mirror agrees with direct
{\it numerical} computations of the superradiant instability
spectrum which characterizes the spinning black-hole-mirror bomb.
\end{abstract}
\bigskip
\maketitle


\section{Introduction}

Within the framework of classical general relativity, the horizon of
a black hole acts as a one-way membrane which irreversibly absorbs
matter and radiation fields. And yet, the intriguing superradiance
phenomenon, first discussed by Zel`dovich \cite{Zel} more than four
decades ago, allows an orbiting cloud made of bosonic particles to
extract rotational energy from a spinning black hole. In particular,
an incident bosonic (integer-spin) field of azimuthal harmonic index
$m$ can be amplified (that is, can gain rotational energy and
angular momentum) as it scatters off a spinning Kerr black hole.
This superradiant scattering phenomenon occurs in the black-hole
spacetime if the proper frequency of the orbiting bosonic field lies
in the bounded regime \cite{Zel,PressTeu2,Viln,Noteun}
\begin{equation}\label{Eq1}
0<\omega<m\Omega_{\text{H}}\  ,
\end{equation}
where $\Omega_{\text{H}}$ is the black-hole angular velocity [see
Eq. (\ref{Eq6}) below].

Interestingly, the energy and angular momentum extraction rates from
the spinning Kerr black hole can grow exponentially in time if the
co-rotating bosonic cloud is prevented from radiating its (growing)
energy to infinity. In particular, Press and Teukolsky
\cite{PressTeu2} have explicitly shown that, by placing the
black-hole-bosonic-field system inside a closed cavity (a reflecting
mirror), one can build a powerful explosive device which
continuously extracts energy and angular momentum from the spinning
black hole \cite{Noteaa,Notemas,HerR}. This composed
Kerr-black-hole-bosonic-field-mirror system is known as the spinning
{\it black-hole bomb} \cite{PressTeu2,Notechg,Bekch,CBHB}.

In a very interesting work, Cardoso et. al. \cite{CarDias} have used
a combination of numerical and analytical techniques to explore the
physical properties of the composed
Kerr-black-hole-scalar-field-mirror system. In particular, the
numerical results presented in \cite{CarDias} have revealed the
interesting fact that, in the linearized regime \cite{Notelin}, the
superradiant instability shuts down if the reflecting mirror, which
is used to confine the Kerr-black-hole-scalar-field system, is
placed too close to the horizon of the spinning black hole.

The critical (minimum) radius of the reflecting mirror,
$r^{\text{c}}_{\text{m}}$, marks the onset of the superradiant
instabilities in the composed Kerr-black-hole-scalar-field-mirror
system \cite{CarDias,BHBM,Hod14w}. In particular, this critical
mirror radius corresponds to {\it stationary} (marginally stable)
confined field modes, which are characterized by the critical
(threshold) frequency
\begin{equation}\label{Eq2}
\omega_{\text{c}}=m\Omega_{\text{H}}\
\end{equation}
for the superradiant scattering phenomenon of bosonic fields in the
spinning Kerr black-hole spacetime.

The physical significance of the critical mirror radius
$r^{\text{c}}_{\text{m}}$ stems from the fact that it is the
smallest (innermost) radius of the reflecting mirror which allows
one to extract rotational energy from the spinning Kerr black hole.
In particular, composed Kerr-black-hole-mirror systems with
$r_{\text{m}}<r^{\text{c}}_{\text{m}}$ can only support stable
(decaying in time) bosonic field configuration, whereas composed
Kerr-black-hole-mirror systems with
$r_{\text{m}}>r^{\text{c}}_{\text{m}}$ can support explosive
(exponentially growing in time) bosonic field configuration.

The main goal of the present paper is to study {\it analytically}
the superradiant instability regime of the composed
Kerr-black-hole-scalar-field-mirror system (the spinning black-hole
bomb of Press and Teukolsky \cite{PressTeu2}). In particular, we
would like to provide a rigorous analytical proof for the existence
of a critical (minimum) radius $r^{\text{c}}_{\text{m}}$ for the
reflecting mirror, below which there is no extraction of rotational
energy from the spinning black hole. Interestingly, we shall show
below that one can use analytical techniques in order to derive an
explicit lower bound [see Eq. ({\ref{Eq37}) below] on the critical
radius $r^{\text{c}}_{\text{m}}$ of the confining mirror which marks
the onset of the exponentially growing superradiant instabilities in
the composed Kerr-black-hole-scalar-field-mirror system.

\section{Description of the system}

We shall analyze the dynamics of a massless scalar field $\Psi$
which is linearly coupled to a spinning Kerr black hole. The curved
black-hole spacetime is described by the line element
\cite{Chan,Kerr,Notebl}
\begin{eqnarray}\label{Eq3}
ds^2=-{{\Delta}\over{\rho^2}}(dt-a\sin^2\theta
d\phi)^2+{{\rho^2}\over{\Delta}}dr^2+\rho^2
d\theta^2+{{\sin^2\theta}\over{\rho^2}}\big[a
dt-(r^2+a^2)d\phi\big]^2\  ,
\end{eqnarray}
where
\begin{equation}\label{Eq4}
\Delta\equiv r^2-2Mr+a^2\ \ \ ; \ \ \ \rho^2\equiv
r^2+a^2\cos^2\theta\  .
\end{equation}
Here $M$ and $a$ are the mass and angular momentum per unit mass of
the Kerr black hole, respectively. The zeros of $\Delta$,
\begin{equation}\label{Eq5}
r_{\pm}=M\pm\sqrt{M^2-a^2}\  ,
\end{equation}
determine the horizon radii of the spinning Kerr black hole. The
black-hole angular velocity is given by \cite{Chan,Kerr}
\begin{equation}\label{Eq6}
\Omega_{\text{H}}={{a}\over{r^2_++a^2}}\  .
\end{equation}

The dynamics of a linearized massless scalar field $\Psi$ in the
black-hole spacetime is described by the familiar Klein-Gordon wave
equation \cite{Teuk}
\begin{equation}\label{Eq7}
\nabla^{\nu}\nabla_{\nu}\Psi=0\  .
\end{equation}
It is convenient to write the scalar eigenfunction $\Psi$ in the
form
\begin{equation}\label{Eq8}
\Psi(t,r;\omega,\theta,\phi)=\sum_{l,m}e^{im\phi}{S_{lm}}(\theta;m,a\omega)
{R_{lm}}(r;M,a,\omega)e^{-i\omega t}\  ,
\end{equation}
where $\omega, l$, and $m$ are respectively the conserved frequency
of the field mode and its angular (spheroidal and azimuthal)
harmonic indices.

Substituting the field decomposition (\ref{Eq8}) into the
Klein-Gordon wave equation (\ref{Eq7}), one finds that the angular
eigenfunctions $S_{lm}$ satisfy the characteristic angular equation
\cite{Teuk,Stro,Heun,Fiz1,Abram,Hodasy}
\begin{eqnarray}\label{Eq9}
{1\over {\sin\theta}}{{d}\over{d\theta}}\Big(\sin\theta {{d
S_{lm}}\over{d\theta}}\Big)
+\Big(K_{lm}-a^2\omega^2\sin^2\theta-{{m^2}\over{\sin^2\theta}}\Big)S_{lm}=0\
.
\end{eqnarray}
Assuming a regular behavior of the eigenfunctions $S_{lm}(\theta)$
at the two poles $\theta=0$ and $\theta=\pi$, one finds that these
angular functions are characterized by a discrete family
$\{K_{lm}\}$ of angular eigenvalues (see \cite{Barma,Hodpp} and
references therein). Interestingly, it was proved in \cite{Barma}
that the angular eigenvalues $\{K_{lm}\}$ associated with the
angular eigenfunctions $\{S_{lm}\}$ are bounded from below by the
characteristic inequality \cite{Notesi}
\begin{equation}\label{Eq10}
K_{lm}\geq m^2+a^2\omega^2\  .
\end{equation}

Substituting the scalar field decomposition (\ref{Eq8}) into the
Klein-Gordon wave equation (\ref{Eq7}), one finds that the radial
eigenfunctions $R_{lm}$ are determined by the differential equation
\cite{Teuk,Stro,Notecoup}
\begin{equation}\label{Eq11}
\Delta{{d} \over{dr}}\Big(\Delta{{d R_{lm}
}\over{dr}}\Big)+\Big\{[\omega(r^2+a^2)-ma]^2
+\Delta(2ma\omega-K_{lm})\Big\}R_{lm}=0\ .
\end{equation}
The ordinary differential equation (\ref{Eq11}), which determines
the radial behavior of the scalar eigenfunctions $R_{lm}(r)$, should
be supplemented by the physical boundary condition of purely ingoing
waves (as measured by a comoving observer) at the horizon of the
spinning black hole \cite{Notemas,HerR,Noteom}:
\begin{equation}\label{Eq12}
R \sim e^{-i(\omega-m\Omega_{\text{H}})y}\ \ \ \text{ for }\ \ \
r\rightarrow r_+\ \ (y\rightarrow -\infty)\ ,
\end{equation}
where the radial coordinate $y$ is determined by the differential
relation $dy=(r^2/\Delta)dr$ [see Eq. (\ref{Eq15}) below]. In
addition, the presence of the reflecting mirror which surrounds the
composed Kerr-black-hole-scalar-field system imposes the boundary
condition \cite{PressTeu2,CarDias,BHBM,Hod14w}
\begin{equation}\label{Eq13}
R(r=r_{\text{m}})=0\
\end{equation}
on the spatial behavior of the confined scalar field modes, where
$r_{\text{m}}$ is the characteristic radius of the confining mirror.

It is worth emphasizing that the sign of $\Im\omega$ in (\ref{Eq8})
determines the (in)stability properties of the confined scalar field
modes. In particular, stable modes (modes decaying exponentially in
time) are characterized by $\Im\omega<0$, whereas superradiantly
unstable modes (modes growing exponentially in time) are
characterized by $\Im\omega>0$. The boundary between stable and
unstable solutions of the composed
Kerr-black-hole-scalar-field-mirror system is marked by the presence
of stationary (marginally stable, with $\Im\omega=0$) field modes,
which are characterized by the critical (threshold) frequency [see
Eq. (\ref{Eq2})]
\begin{equation}\label{Eq14}
\omega_{\text{c}}=m\Omega_{\text{H}}\
\end{equation}
for the superradiant scattering phenomenon in the spinning Kerr
black-hole spacetime.

\section{The effective radial potential of the composed Kerr-black-hole-scalar-field system}

The differential equation (\ref{Eq11}) for the radial scalar
eigenfunctions, together with the boundary conditions (\ref{Eq12})
and (\ref{Eq13}), determine a discrete set of complex resonant
frequencies $\{\omega_n(r_{\text{m}};M,a)\}$ which characterize the
composed Kerr-black-hole-scalar-field-mirror system
\cite{PressTeu2,CarDias,BHBM,Hod14w,Noteqnm,QNMK}. As discussed
above, Cardoso et. al. \cite{CarDias} have performed a very
interesting numerical study of the complex resonant spectrum which
characterizes the spinning black-hole bomb. In particular, Cardoso
et. al. \cite{CarDias} have revealed the interesting fact that the
superradiant instability shuts down if the reflecting mirror is
placed too close to the horizon of the spinning black hole.

The main goal of the present paper is to explore the physical
properties of the marginally stable (stationary) confined field
configurations which characterize the composed
Kerr-black-hole-scalar-field-mirror system. In particular, we would
like to provide a rigorous analytical proof for the existence of a
critical (minimum) radius $r=r^{\text{c}}_{\text{m}}$, which
characterizes the reflecting mirror, below which there is no
extraction of rotational energy from the spinning Kerr black hole.
Furthermore, below we shall derive {\it analytically} an explicit
lower bound on the critical radius $r^{\text{c}}_{\text{m}}(M,a)$ of
the confining mirror which marks the onset of the exponentially
growing superradiant instabilities in this composed physical system.

In order to study the physical properties which characterize the
composed Kerr-black-hole-scalar-field-mirror system, we shall first
transform the radial equation (\ref{Eq11}) into the more familiar
form of a Schr\"odinger-like differential equation. Using in Eq.
(\ref{Eq11}) the differential relation \cite{Notemap}
\begin{equation}\label{Eq15}
dy={{r^2}\over{\Delta}}dr\  ,
\end{equation}
one obtains the characteristic Schr\"odinger-like differential
equation
\begin{equation}\label{Eq16}
{{d^2\psi}\over{dy^2}}-V(y)\psi=0\
\end{equation}
for the radial scalar eigenfunction
\begin{equation}\label{Eq17}
\psi=rR\  ,
\end{equation}
where the effective radial potential in (\ref{Eq16}) is given by
\begin{equation}\label{Eq18}
V=V(r;\omega,M,a,l,m)={{2\Delta}\over{r^6}}(Mr-a^2)+{{\Delta}\over{r^4}}
(K_{lm}-2ma\omega)-{{1}\over{r^4}}[\omega(r^2+a^2)-ma]^2\ .
\end{equation}

In the next section we shall study the near-horizon behavior of the
radial eigenfunction $\psi$ which characterizes the spatial
properties of the confined scalar fields in the spinning Kerr
black-hole spacetime. To that end, we shall first explore the
near-horizon properties of the effective radial potential $V(r)$
which governs the dynamics of the linearized scalar fields in the
black-hole spacetime.

\section{The near-horizon spatial behavior of the radial scalar eigenfunctions}

The main goal of the present paper is to study the onset of the
superradiant instabilities in the composed
Kerr-black-hole-scalar-field-mirror system. Thus, we shall
henceforth explore the physical properties of the marginally stable
scalar modes, which are characterized by the critical (threshold)
frequency (\ref{Eq2}) for the superradiant scattering phenomenon in
the spinning Kerr black-hole spacetime. As emphasized above, the
boundary between stable ($\Im\omega<0$) and unstable ($\Im\omega>0$)
composed Kerr-black-hole-scalar-field-mirror configurations is
marked by the presence of these stationary (marginally stable, with
$\Im\omega=0$) confined scalar field modes.

For later purposes, we shall prove in this section that the radial
scalar eigenfunction $\psi$, which characterizes the stationary
(marginally stable) confined scalar field modes of the composed
Kerr-black-hole-scalar-field-mirror system, is a positive
\cite{Notepp}, increasing, and convex function in the near-horizon
\begin{equation}\label{Eq19}
x\ll\tau
\end{equation}
region of the spinning Kerr black-hole spacetime, where
\begin{equation}\label{Eq20}
x\equiv {{r-r_+}\over{r_+}}\ \ \ \ \text{and} \ \ \ \
\tau\equiv{{r_+-r_-}\over{r_+}}\
\end{equation}
are a dimensionless radial coordinate and the dimensionless
black-hole temperature, respectively.

Substituting the resonant oscillation frequency (\ref{Eq2}), which
characterizes the marginally stable (stationary) confined field
modes, into the expression (\ref{Eq18}) for effective radial
potential of the composed black-hole-scalar-field system, one finds
the near-horizon behavior
\begin{equation}\label{Eq21}
r^2_+V(x\to0)=F\tau\cdot x+O(x^2)\  ,
\end{equation}
where the expansion coefficient in (\ref{Eq21}) is given by
\begin{equation}\label{Eq22}
F\equiv K_{lm}-{{2(ma)^2}\over{r^2_++a^2}}+\tau\ .
\end{equation}
Furthermore, substituting the lower bound (\ref{Eq10}) on the
angular eigenvalues into (\ref{Eq22}), one finds the characteristic
inequality
\begin{equation}\label{Eq23}
F>m^2\Big({{r_+}\over{r_++r_-}}\Big)^2+\tau>0\
\end{equation}
for the near-horizon expansion coefficient of the effective radial
potential. We therefore conclude that, in the near-horizon region
(\ref{Eq19}), the effective radial potential (\ref{Eq18}) which
characterizes the stationary (marginally stable)
Kerr-black-hole-scalar-field-mirror configurations has the form of
an effective potential barrier with [see Eqs. (\ref{Eq21}) and
(\ref{Eq23})]
\begin{equation}\label{Eq24}
V\geq0\ \ \ \ \text{for} \ \ \ \ x\ll\tau\  .
\end{equation}

Taking cognizance of the differential relation (\ref{Eq15}), one
finds the near-horizon ($x\ll\tau$) relation
\begin{equation}\label{Eq25}
y={{r_+}\over{\tau}}\ln(x)+O(x)\  ,
\end{equation}
which implies \cite{Noteyas}
\begin{equation}\label{Eq26}
x=e^{\tau y/r_+}[1+O(e^{\tau y/r_+})].
\end{equation}
Substituting (\ref{Eq21}) and (\ref{Eq26}) into (\ref{Eq16}), one
obtains the Schr\"odinger-like differential equation
\begin{equation}\label{Eq27}
{{d^2\psi}\over{d\tilde y^2}}-{{4F}\over{\tau}}e^{2\tilde y}\psi=0\
\ \ \ \text{with} \ \ \ \ \tilde y\equiv {{\tau}\over{2r_+}}y\
\end{equation}
in the near-horizon region (\ref{Eq19}). The physically acceptable
solution [that is, the solution which respects the physical boundary
condition (\ref{Eq12}) at the black-hole horizon] of the
Schr\"odinger-like radial equation (\ref{Eq27}) is given by
\cite{Noteab1}
\begin{equation}\label{Eq28}
\psi(y)=I_0\Big(2\sqrt{{{F}\over{\tau}}}e^{\tau y/2r_+}\Big)\  .
\end{equation}
Using the well-known mathematical properties of the modified Bessel
function of the first kind $I_0$ \cite{Abram}, one finds from the
radial solution (\ref{Eq28}) that the scalar eigenfunction $\psi$,
which characterizes the spatial behavior of the stationary
(marginally stable) Kerr-black-hole-scalar-field-mirror
configurations, is a positive, increasing, and convex function in
the near-horizon region (\ref{Eq19}). That is,
\begin{equation}\label{Eq29}
\{\psi>0\ \ \ ;\ \ \ {{d\psi}\over{dy}}>0\ \ \ ;\ \ \
{{d^2\psi}\over{dy^2}}>0\}\ \ \ \ \text{for}\ \ \ \ 0<x\ll\tau\ .
\end{equation}

Taking cognizance of Eqs. (\ref{Eq13}) and (\ref{Eq29}), one
concludes that the radial scalar eigenfunction $\psi$, which
characterizes the stationary confined scalar field modes, must have
(at least) one maximum point, $x=x_{\text{max}}$, between the
horizon of the spinning Kerr black hole [where $\psi$ is a positive
and increasing function, see (\ref{Eq29})] and the radial location
$r=r_{\text{m}}$ of the reflecting mirror [where $\psi$ vanishes,
see (\ref{Eq13})]. For later purposes, it is important to stress the
fact that, at the maximum point $x=x_{\text{max}}$, the scalar
eigenfunction $\psi$ is characterized by the inequalities
\begin{equation}\label{Eq30}
\{\psi>0\ \ \ \text{and}\ \ \ {{d^2\psi}\over{dy^2}}<0\}\ \ \
\text{for}\ \ \ x=x_{\text{max}}\  .
\end{equation}

\section{The superradiant instability regime of the composed Kerr-black-hole-scalar-field-mirror system}

The analysis presented in the previous section has revealed the
important fact that the eigenfunction $\psi$, which characterizes
the radial behavior of the stationary (marginally stable) confined
scalar field modes in the spinning Kerr black-hole spacetime, must
have (at least) one maximum point, $r=r_{\text{max}}$, within the
interval
\begin{equation}\label{Eq31}
r_+<r_{\text{max}}<r_{\text{m}}\  .
\end{equation}
In particular, taking cognizance of Eqs. (\ref{Eq16}) and
(\ref{Eq30}), one realizes that the radial potential (\ref{Eq18}),
which governs the dynamics of the scalar field in the spinning Kerr
black-hole spacetime, must have a negative value at this maximum
point. That is,
\begin{equation}\label{Eq32}
V(r=r_{\text{max}})<0\  .
\end{equation}
As we shall now show, one can use this characteristic property of
the effective radial potential in order to derive a generic lower
bound on the critical radius $r^{\text{c}}_{\text{m}}(M,a)$ of the
confining mirror which marks the onset of the superradiant
instabilities in the composed Kerr-black-hole-scalar-field-mirror
system.

Substituting the critical oscillation frequency (\ref{Eq2}) of the
stationary (marginally stable) confined field modes into
(\ref{Eq18}), and using the characteristic lower bound (\ref{Eq10})
on the angular eigenvalues, one finds that the effective radial
potential of the composed black-hole-field-mirror system is
characterized by the inequality
\begin{equation}\label{Eq33}
V(r;\omega=\omega_{\text{c}})\geq m^2\cdot
{{r-r_+}\over{r^3(r^2_++a^2)^2}}(-a^2r^2-a^2r_+r+2Mr^3_+)+{{2\Delta}\over{r^6}}(Mr-a^2)\
.
\end{equation}
Furthermore, using the relation $Mr-a^2\geq0$ \cite{Noteineq}, one
finds from (\ref{Eq33}) the characteristic inequality
\begin{equation}\label{Eq34}
V(r;\omega=\omega_{\text{c}})\geq m^2\cdot
{{r-r_+}\over{r^3(r^2_++a^2)^2}}(-a^2r^2-a^2r_+r+2Mr^3_+)\
\end{equation}
for the effective radial potential.

The inequalities (\ref{Eq32}) and (\ref{Eq34}) imply that the
maximum point $r=r_{\text{max}}$ of the radial scalar eigenfunction
$\psi$ is characterized by the inequality
\begin{equation}\label{Eq35}
-a^2r^2_{\text{max}}-a^2r_+r_{\text{max}}+2Mr^3_+<0\  ,
\end{equation}
which yields the simple lower bound
\begin{equation}\label{Eq36}
r_{\text{max}}>{{r_+}\over
{2}}\Big(\sqrt{1+{{8Mr_+}\over{a^2}}}-1\Big)\  .
\end{equation}
Finally, taking cognizance of the inequalities (\ref{Eq31}) and
(\ref{Eq36}), one obtains the lower bound \cite{Notebdb}
\begin{equation}\label{Eq37}
r^{\text{c}}_{\text{m}}>{{r_+}\over
{2}}\Big(\sqrt{1+{{8Mr_+}\over{a^2}}}-1\Big)
\end{equation}
on the critical radius $r^{\text{c}}_{\text{m}}(M,a)$ of the
reflecting mirror which supports the stationary (marginally stable)
confined field configurations in the Kerr black-hole spacetime.

Interestingly, the expression on the r.h.s of (\ref{Eq37}) is a
decreasing function of the black-hole rotation parameter $a$. In
particular, in the near-extremal $a/M\to1$ limit one finds the
simple lower bound [see Eqs. (\ref{Eq20}) and (\ref{Eq37})]
\begin{equation}\label{Eq38}
x^{\text{c}}_{\text{m}}>{\tau\over 3}\
\end{equation}
on the dimensionless critical radius
$x^{\text{c}}_{\text{m}}=(r^{\text{c}}_{\text{m}}-r_+)/r_+$ of the
confining mirror.

We would like to stress the fact that the physical significance of
the analytically derived lower bound (\ref{Eq37}) on the critical
radius $r^{\text{c}}_{\text{m}}(M,a)$ \cite{Notebdb} of the
confining mirror stems from the fact that this inequality provides a
necessary condition for the development of the exponentially growing
superradiant instabilities in the composed
Kerr-black-hole-scalar-field-mirror system (the spinning
black-hole-mirror bomb of Press and Teukolsky \cite{PressTeu2}).

\section{Numerical confirmation}

It is of physical interest to verify the validity of the
analytically derived lower bounds (\ref{Eq37}) and (\ref{Eq38}) on
the superradiant instability regime of the composed
Kerr-black-hole-scalar-field-mirror system (the spinning
black-hole-mirror bomb). The resonant oscillation spectrum of this
composed physical system was investigated numerically in
\cite{CarDias,BHBM,Hod14w}. In Table \ref{Table1} we present the
dimensionless ratio
$x^{\text{stat}}_{\text{m}}/x^{\text{bound}}_{\text{m}}$ for
near-extremal black holes and for various values of the azimuthal
harmonic index $m$ which characterizes the confined scalar mode,
where $x^{\text{stat}}_{\text{m}}$ is the {\it numerically} computed
\cite{Hod14w} value of the dimensionless mirror radius which
corresponds to the stationary (marginally stable) confined scalar
fields \cite{Notesbo}, and $x^{\text{bound}}_{\text{m}}$ is the {\it
analytically} derived lower bound (\ref{Eq38}) on the dimensionless
critical radius of the confining mirror. One finds from Table
\ref{Table1} that the superradiant instability regime of the
composed Kerr-black-hole-scalar-field-mirror system is characterized
by the inequality
$x^{\text{stat}}_{\text{m}}/x^{\text{bound}}_{\text{m}}>1$, in
agreement with the analytically derived lower bound (\ref{Eq38}).

\begin{table}[htbp]
\centering
\begin{tabular}{|c|c|c|c|c|c|c|}
\hline $l=m$ & \ \ 5 \ \ & \ \ 25 \ \ & \ \ 50 \ \ & \ \ 100 \ \ & \ \ 150 \ \ & \ \ 200 \ \ \\
\hline \ \ $x^{\text{stat}}_{\text{m}}/x^{\text{bound}}_{\text{m}}$
\ \ &\ \ \ 10.35\ \ \ \ &\ \ \ 2.25\ \ \ \ &\ \ \ 1.65\ \ \ \ &\ \ \
1.38\ \ \ \ &\ \ \
1.29\ \ \ \ &\ \ \ 1.23\ \ \ \ \\
\hline
\end{tabular}
\caption{The superradiant instability regime of the composed
Kerr-black-hole-scalar-field-mirror system (the Press-Teukolsky
spinning black-hole-mirror bomb \cite{PressTeu2}). We display the
dimensionless ratio
$x^{\text{stat}}_{\text{m}}/x^{\text{bound}}_{\text{m}}$ for
near-extremal black holes and for various values of the azimuthal
harmonic index $m$ which characterizes the confined scalar mode,
where $x^{\text{stat}}_{\text{m}}$ is the {\it numerically} computed
\cite{Hod14w} value of the dimensionless mirror radius which
corresponds to the stationary (marginally stable) confined scalar
fields \cite{Notesbo}, and $x^{\text{bound}}_{\text{m}}$ is the {\it
analytically} derived lower bound (\ref{Eq38}) on the dimensionless
critical radius of the confining mirror. One finds that the composed
Kerr-black-hole-scalar-field-mirror system is characterized by the
inequality
$x^{\text{stat}}_{\text{m}}/x^{\text{bound}}_{\text{m}}>1$, in
agreement with the analytically derived lower bound (\ref{Eq38}).}
\label{Table1}
\end{table}

\section{The case of a confined massive scalar field}

Our analysis can be extended to the physically interesting case of
confined {\it massive} scalar fields linearly coupled to a spinning
Kerr black hole. In this case one finds that Eq. (\ref{Eq35}) should
be replaced by \cite{Notemas}
\begin{equation}\label{Eq39}
\alpha\cdot a^2r^2_{\text{max}}-a^2(2M+\alpha\cdot
r_-)r_{\text{max}}+2Mr^3_+<0\ ,
\end{equation}
where \cite{Noteaph}
\begin{equation}\label{Eq40}
\alpha\equiv
{{\mu^2-\omega^2_{\text{c}}}\over{\omega^2_{\text{c}}}}\  .
\end{equation}
The inequality (\ref{Eq39}) yields the lower bound [see
(\ref{Eq31})]
\begin{equation}\label{Eq41}
r^{\text{c}}_{\text{m}}>{{2M+\alpha r_- -\sqrt{(2M+\alpha
r_-)^2-8\alpha Mr^3_+/a^2}}\over{2\alpha}}\
\end{equation}
on the critical radius $r^{\text{c}}_{\text{m}}(M,a,\mu)$ of the
reflecting mirror which supports the stationary (marginally stable)
massive scalar field configuration in the spinning Kerr black-hole
spacetime.

It is worth emphasizing the fact that the inequality (\ref{Eq39}),
which is a necessary condition for the development of the
exponentially growing superradiant instabilities in the
Press-Teukolsky spinning black-hole-mirror bomb \cite{PressTeu2},
can only be satisfied in the restricted regime \cite{Notemas}
\begin{equation}\label{Eq42}
\alpha<
{{2M\big[2r^2_+-r^2_--2r_+({r^2_+-r^2_-})^{1/2}\big]}\over{r^3_-}}\
.
\end{equation}
We therefore conclude that confined massive scalar fields whose
masses lie outside the regime (\ref{Eq42}) {\it cannot} extract the
rotational energy of the corresponding spinning Kerr black hole.

\section{Summary and discussion}

The superradiant instability regime of the spinning black-hole bomb
\cite{PressTeu2,CarDias,BHBM,Hod14w} was studied {\it analytically}.
This physical system, first designed by Press and Teukolsky
\cite{PressTeu2}, is composed of a spinning Kerr black hole of
horizon radii $\{r_+,r_-\}$ which is surrounded by a reflecting
mirror of radius $r_{\text{m}}$. Thanks to the intriguing
superradiant amplification mechanism \cite{Zel,PressTeu2,Viln}, an
orbiting scalar field, which is placed between the black-hole
horizon and the reflecting mirror, is able to extract rotational
energy and angular momentum from the spinning Kerr black hole. The
role of the confining mirror is to prevent the amplified scalar
field from radiating this extracted energy to infinity. As a
consequence, it is well known that this composed
Kerr-black-hole-scalar-field-mirror system may develop exponentially
growing superradiant instabilities
\cite{PressTeu2,CarDias,BHBM,Hod14w}.

In the present paper we have addressed the following physically
interesting question: Is it possible to extract the rotational
energy of a spinning Kerr black hole by placing a reflecting mirror
(whose role is to confine the superradiantly amplified scalar
fields) arbitrarily close to the black-hole horizon? An analytical
answer to this important question would certainly enrich our
understanding of the intriguing phenomena of superradiant
amplification of bosonic fields in rotating black-hole spacetimes.

Moreover, it is worth noting that the answer to this intriguing
question may one day be of practical importance \cite{Hod14w}.
Astrophysical observations made in recent years \cite{Astro} have
provided compelling evidence that spinning black holes are
ubiquitous in our Universe. Thus, Kerr black holes may serve our
civilization in the future as enormous sources of clean (and
probably cheap) energy. In this physical process of energy
extraction from a spinning black hole, one would like to build the
reflecting mirror as close as possible to the horizon of the central
black hole in order to save construction materials. It is therefore
of practical importance to determine the minimum radius of the
reflecting mirror which allows the extraction of rotational energy
from the spinning Kerr black hole.

The main results derived in this paper and their physical
implications are as follows:

(1) Using analytical techniques, we have proved that, in order for
the exponentially growing superradiant instabilities to develop in
this composed black-hole-field-mirror system, the dimensionless
radius of the reflecting mirror must be bounded from below by the
relation [see Eq. (\ref{Eq37})] \cite{Notebdb,Noterar}
\begin{equation}\label{Eq43}
{{r^{\text{c}}_{\text{m}}}\over{r_+}}>{1\over
2}\Big(\sqrt{1+{{8M}\over{r_-}}}-1\Big)\  .
\end{equation}
It is worth stressing the fact that the physical significance of
this analytically derived lower bound on the dimensionless critical
radius of the confining mirror stems from the fact that the compact
relation (\ref{Eq39}) provides a necessary condition for the
development of the exponentially growing superradiant instabilities
in the Press-Teukolsky spinning black-hole-mirror bomb
\cite{PressTeu2}.

(2) In a very interesting work, Cardoso et. al. \cite{CarDias} have
studied the instability properties of the spinning
black-hole-scalar-field-mirror bomb. In particular, the numerical
results presented in \cite{CarDias} have revealed the interesting
fact that the critical (innermost) radius $r^{\text{c}}_{\text{m}}$
of the confining mirror \cite{Notebdb} is a {\it decreasing}
function of both the black-hole angular momentum $a$ and the
azimuthal harmonic index $m$ of the confined field mode. It is
important to note, however, that the analytically derived lower
bound (\ref{Eq43}) reveals the fact that, even in the double
asymptotic limit $a/M\to1$ with $m\to\infty$, the reflecting mirror
{\it cannot} be placed arbitrarily close to the black-hole horizon.
In particular, in the extremal $a/M\to 1$ ($\tau\to0$) limit one
finds the lower bound [see Eq. (\ref{Eq38})]
\begin{equation}\label{Eq44}
{{x^{\text{c}}_{\text{m}}}\over{\tau}}>{1\over 3}\
\end{equation}
on the dimensionless critical radius of the reflecting mirror.

(3) It is worth noting that the numerical results presented in
\cite{Hod14w} predict the asymptotic value
${{x^{\text{c}}_{\text{m}}}\over{\tau}}\simeq0.36$ for the
dimensionless critical radius of the reflecting mirror in the double
asymptotic limit $a/M\to1$ with $m\to\infty$. This {\it numerically}
computed asymptotic value is consistent with (and, in fact, quite
close to) the {\it analytically} derived lower bound
${{x^{\text{c}}_{\text{m}}}\over{\tau}}>1/3$ [see Eq. (\ref{Eq44})]
on the dimensionless critical radius of the reflecting mirror. Thus,
our results [and, in particular, the analytically derived lower
bounds (\ref{Eq43}) and (\ref{Eq44})] provide a quantitative
analytical explanation for the numerical results presented in
\cite{Hod14w} for the asymptotic large-$m$ behavior of the composed
Kerr-black-hole-scalar-field-mirror system (the Press-Teukolsky
spinning black-hole-mirror bomb).

(4) It is important to stress the fact that the existence of a lower
bound [see the analytically derived relation (\ref{Eq43})] on the
radius of the confining mirror is a highly non-trivial feature of
the Kerr black-hole spacetime. In particular, it is physically
interesting to contrast the physical properties of the {\it
spinning} Kerr-black-hole-scalar-field-mirror system with the
corresponding properties of the {\it charged}
Reissner-Nordstr\"om-black-hole-scalar-field-mirror system studied
in \cite{CBHB}. Our analytically derived results have revealed the
fact that in the former case the radius of the confining mirror is
bounded from below by (\ref{Eq43}) (that is, the confining mirror
cannot be placed arbitrarily close to the horizon of the spinning
Kerr black hole), whereas in the later case the confining mirror can
be placed arbitrarily close to the horizon of the charged
Reissner-Nordstr\"om black hole \cite{CBHB}.

(5) We have also generalized the analysis to the physically
interesting case of confined {\it massive} scalar fields linearly
coupled to a spinning Kerr black hole. In particular, we have
explicitly proved that confined massive scalar fields whose masses
lie outside the regime (\ref{Eq42}) {\it cannot} extract the
rotational energy of the corresponding spinning Kerr black hole.

Finally, we would like to stress the fact that, in the present
analysis, the confined scalar fields were treated at the
perturbative (linearized) level. The analytical results derived in
this paper are therefore expected to be valid in the ignition stage
of the superradiant instabilities, when the amplitude of the
confined field is still small and its dynamics in the black-hole
spacetime can still be treated at the linearized perturbative level.
As we explicitly shown in this paper, the physical properties of the
Press-Teukolsky spinning black-hole bomb (the composed
Kerr-black-hole-scalar-field-mirror system \cite{PressTeu2}) can be
studied {\it analytically} within the framework of this perturbative
(linearized) approach. We believe that it would be highly
interesting (and physically important) to use more sophisticated
{\it numerical} techniques in order to explore the late-time
non-linear development of the explosive superradiant instabilities
in this composed Kerr-black-hole-scalar-field-mirror system.

\bigskip
\noindent
{\bf ACKNOWLEDGMENTS}
\bigskip

This research is supported by the Carmel Science Foundation. I thank
Yael Oren, Arbel M. Ongo, Ayelet B. Lata, and Alona B. Tea for
stimulating discussions.



\begin{thebibliography}{99}
\bibitem{Noteun} We shall use natural units in which $G=c=\hbar=1$.
\bibitem{Noteaa} It is worth mentioning that, for massive fields, there is an alternative
confinement mechanism which is naturally provided by the mutual
gravitational attraction between the Kerr black hole and the
orbiting bosonic field \cite{Notemas,HerR}. In particular, field
modes in the regime $\omega<\mu$ (here $\mu$ is the proper mass of
the scattered field) are prevented from escaping to infinity
\cite{Notemas,HerR}.
\bibitem{Notelin} That is, in the regime of a linearized (test) scalar
field propagating on the fixed background of the composed
Kerr-black-hole-mirror system.
\bibitem{Notebl} The line element (\ref{Eq3}) of the Kerr black-hole spacetime
is expressed in terms of the Boyer-Lindquist coordinates
$(t,r,\theta,\phi)$.
\bibitem{Notesi} Interestingly, it was shown in \cite{Hodpp} that the lower bound (\ref{Eq10})
on the angular eigenvalues which characterize the angular
eigenfunctions can be saturated in the eikonal $l=m\gg1$ limit. In
particular, one finds the simple asymptotic behavior
$K_{mm}=m^2[1+O(m^{-1})]+a^2\omega^2$ \cite{Hodpp}.
\bibitem{Notecoup} It is worth emphasizing the fact that, due to the rotation of the black-hole spacetime,
the differential equation (\ref{Eq11}) for the radial eigenfunctions
$R_{lm}$ is coupled to the differential equation (\ref{Eq9}) for the
angular eigenfunctions $S_{lm}$ by the characteristic angular
eigenvalues $K_{lm}$.
\bibitem{Noteom} We shall henceforth omit the angular harmonic indices $\{l,m\}$ for brevity.
\bibitem{Noteqnm} It is worth emphasizing again that the caged (bounded) scalar field modes
that we study in the present paper are characterized by the boundary
condition (\ref{Eq13}). This boundary condition, which is imposed at
the finite radius $r=r_{\text{m}}$, is dictated by the presence of
the reflecting mirror around the black hole. On the other hand, the
more familiar quasinormal resonant frequencies, which characterize
the linearized dynamics of scalar fields in an asymptotically flat
black-hole spacetime, are characterized by an outgoing boundary
condition at asymptotically large distances $r\to\infty$ from the
black hole, see \cite{QNMK} and references therein.
\bibitem{Notemap} Note that, in the black-hole spacetime, the differential relation (\ref{Eq15}) maps the radial
coordinate $r\in [r_+,\infty]$ into $y\in [-\infty,+\infty]$.
\bibitem{Notepp} One can assume $R(r=r_+)\geq0$ for the radial scalar eigenfunction without loss of generality.
\bibitem{Noteyas} It is worth noting that the near-horizon $x\ll\tau$ region corresponds to $\tau y/r_+\to
-\infty$ [see Eq. (\ref{Eq25})]. This implies $e^{\tau y/r_+}\to 0$
in the near-horizon region.
\bibitem{Noteab1} Here we have used equations (9.1.54) and (9.6.3) of \cite{Abram}.
\bibitem{Noteineq} Note that this inequality follows from the series of inequalities $Mr-a^2\geq
Mr_+-a^2=r_+(M-r_-)\geq0$ in the external region $r\geq r_+$ of the
spinning Kerr black-hole spacetime.
\bibitem{Noteaph} Note that the massless scalar field case
corresponds to $\alpha=-1$.
\bibitem{Noterar} Here we have used the relation $a^2=r_+r_-$ for
the spinning Kerr black hole.
\end{thebibliography}
\end{document}